# Earthshine as an Illumination Source at the Moon

David A. Glenar[a,b], Timothy J. Stubbs[c,b], Edward W. Schwieterman[d,e,f,g],
Tyler D. Robinson[h,g], and Timothy A. Livengood[c]

[a] *Center for Space Science and Technology, University of Maryland, Baltimore County, Baltimore, MD., 21218*
[b] *Solar System Exploration Research Virtual Institute/SSERVI, NASA Ames Research Center, Moffett Field, CA, 94035*
[c] *NASA Goddard Space Flight Center, Greenbelt, MD, 20771*
[d] *Department of Earth Sciences, University of California, Riverside, Riverside, CA 92521*
[e] *NASA Postdoctoral Program, Universities Space Research Association, Columbia, Maryland, USA*
[f] *Blue Marble Space Institute of Science, Seattle, Washington, USA*
[g] *NASA Astrobiology Institute Virtual Planetary Laboratory, Seattle, WA 98195*
[h] *Department of Physics and Astronomy, Northern Arizona University, Flagstaff, AZ, 86011*

**Abstract**

Earthshine is the dominant source of natural illumination on the surface of the Moon during lunar night, and at some locations within permanently shadowed regions (PSRs) near the poles that never receive direct sunlight. As such, earthshine has the potential to enable the scientific investigation and exploration of conditions in areas of the Moon that are either temporarily or permanently hidden from the Sun. Under certain circumstances, the heat flux from earthshine could also influence the transport and cold-trapping of volatiles present in the very coldest areas within PSRs. In this study, Earth's spectral irradiance, as it would appear at the Moon in the solar reflectance band (0.3–3.0 μm) and at thermal emission wavelengths (3–50 μm), is examined with a suite of model image cubes and whole-disk spectra created using the Virtual Planetary Laboratory (VPL) three-dimensional (latitude, longitude and altitude) modeling capability. At the Moon, the broadband, hemispherical irradiance from Earth at full-phase is approximately 0.15 W m$^{-2}$ with comparable contributions from solar reflectance and thermal emission; for context, this about 0.01% that of solar irradiance and has an equivalent temperature of around 40 K. Over the simulated timeframe, spanning two lunations, Earth's thermal irradiance shows very little net change (less than a few mWm$^{-2}$ resulting from cloud variability and the south-to-north motion of the sub-observer latitude on Earth). In the solar band, Earth's diurnally averaged light curve at phase angles $g \leq 60°$ is well-fit using a Henyey-Greenstein integral phase function. At wavelengths longward of about 0.7 μm, near the well-known vegetation "red edge", Earth's reflected solar radiance shows significant diurnal modulation as a result of the broad maximum in projected landmass at terrestrial longitudes between 60°W and 0°, as well as from the distribution of clouds. A simple formulation with adjustable coefficients is presented, condensed from the VPL model grid, for estimating Earth's hemispherical irradiance at the Moon as a function of wavelength, phase angle and sub-observer coordinates (terrestrial latitude and longitude). Uncertainties in any one prediction are estimated to be 10–12% at 0.3 μm, rising to >25% near 2.5 μm as a result of the increasing relative brightness and unpredictable influence of clouds. Although coefficient values are derived from a suite of spring equinox models, the approximation appears to be valid for all seasons, to within the stated uncertainties. It is demonstrated that earthshine is sufficiently bright to serve as a natural illumination source for optical measurements on a robotic lander/rover, allowing the identification of water ice mixed with regolith at the percent-level of fractional area.

# 1. Introduction

Earthlight refers to the combined irradiance from reflected sunlight and thermal emission from Earth, which provides an illumination source for other objects in space. The earthlight that illuminates the Moon is known as earthshine, the effects of which are most easily seen by Earth-based observers around New Moon. Averaged over a lunation, the irradiance from earthshine is dwarfed by that from incident sunlight; however, during lunar night it is by far the dominant source of natural illumination. As demonstrated here, the broadband earthshine irradiance approaches 150 mW m$^{-2}$ (about ten thousandth that from the Sun) near zero phase with comparable contributions from solar reflectance and thermal emission. Earthshine can also be dominant in areas within lunar polar craters that are never exposed to direct sunlight, referred to as permanently shadowed regions (PSRs). Although not directly illuminated by the Sun, PSRs can be exposed to sunlight scattered from surrounding terrain with average irradiances ranging from ~<1–2000 mW m$^{-2}$ (*Mazarico et al., 2011*). Near the poles of the Moon, the elevation angle of Earth periodically exceeds 6° due to lunar latitude libration. As a result, earthshine illuminates equatorward facing walls of some PSRs, as well as portions of high latitude terrain with local slopes tilted toward Earth (*Mazarico et al., 2011; Mazarico and Nicholas, 2016; Glenar et al., 2016*). Therefore, it is likely that many areas exist within PSRs that are typically better illuminated by earthshine than by scattered sunlight. The brightness of earthshine makes it a promising natural illumination source for scientific investigations and exploration of lunar environments and terrains both on the night side and in and around PSRs, especially in areas not well-lit by scattered sunlight. This is ably illustrated by images of zodiacal light taken using the Clementine star trackers that also clearly show the lunar surface being well-illuminated by earthshine (*Hahn et al. 2002; Glenar et al. 2014*). Future optical instruments on polar surface missions could utilize earthshine as a broadband illumination source for imaging and/or spectroscopy, either by itself or as a complement to active illumination.

Earthshine may also make a non-negligible contribution to the thermal environment in localized areas of PSRs where scattered sunlight is relatively weak or absent, and when Earth is visible above the local horizon. Such conditions will be more widespread during polar winter. Under these circumstances, the most significant terms in determining thermal balance would be earthshine and the Moon's internal heat flux. The average lunar mantle heat flux is currently estimated to be 9–13 mW m$^{-2}$ (*Siegler and Smrekar, 2014*), which is over an order-of-magnitude smaller than earthshine at its most intense. Therefore, earthshine could thermally influence some of the coldest locations on the Moon. This raises the possibility that earthshine plays a role in the distribution and transport of some volatile atomic and simple organic species in PSRs that are only thermally stable at a few tens of kelvin (Zhang and Paige, 2009, 2010; Paige et al., 2010). There may even be instances where earthshine influences the stability of water ice, by marginally increasing surface temperatures ($\Delta T < 1$ K) at locations that are already warmed by scattered sunlight to nearly the sublimation temperature of ~110K. Any influence earthshine might have on water ice in PSRs would likely be very subtle.

Models for Earth's spectral radiance have gradually improved in accuracy, motivated by the need to understand climate change, as well as the prospects for future Earth-like extrasolar planet detection. In the past, models of Earth's albedo have been strongly guided by ground-based measurements of earthshine on the Moon, aided by near simultaneous measurements of cloud cover by Earth-orbiting satellites (*Goode et al., 2001; Qiu et al., 2003; Palle et al., 2003, 2004; Palle et al., 2009*). Spectral measurements (*Woolf et al., 2002; Turnbull et al., 2006*) reveal the major features in Earth's visible and near-IR reflectance spectrum. Ground-based earthshine measurements have the limitation that only a fraction of the Earth's Moon-facing hemisphere can be measured in a single observation. This difficulty is compounded by the Moon's reflectance function, which is direction dependent. Increasingly accurate Earth radiance models have been developed using data from Earth-observing satellites and occasional flybys of interplanetary probes (*Tinetti et al., 2006a, 2006b; Oakley and Cash, 2009*). One such example is the NASA Astrobiology Institute's Virtual Planetary Laboratory (VPL) three-dimensional (longitude, latitude and vertical) model that utilizes date-specific measurements from multiple Earth-orbiting satellites (*Robinson et al., 2010, 2011; Schwieterman et al., 2015*). VPL model predictions were used to perform an in-flight recalibration of the UV/Vis spectrometer aboard the Lunar Crater Observation and Sensing Satellite (LCROSS) after completing several Earth observations (*Robinson et al. 2014*). In parallel with the evolution of accurate Earth radiance models, the geometry of Earth's illumination on lunar topography has been computed at spatial scales of order 100 m, using digital elevation models (DEMs) derived from measurement by the Lunar Observer Laser Altimeter (LOLA) aboard the Lunar Reconnaissance Orbiter (LRO) (*Mazarico et al., 2011*). By combining these resources, it is possible to quantify Earth's radiative influence within PSRs, and even search for the possible influence of earthshine present in existing data sets.

In this study, we review the major features of earthshine's solar reflectance and thermal spectra using both spatially resolved and disk-integrated VPL models near spring equinox, as described in Sections 2 and 3. In Section 4, the strong diurnal modulation of Earth's near-IR reflectance, as well as the orbital light curves and sources of variability, are discussed and quantified as a function of wavelength and orbital phase. In Section 5, we construct a semi-empirical spring season approximation for Earth's hemispherical irradiance at the Moon, as a function of wavelength, phase and (terrestrial) sub-observer coordinates. The chosen season avoids extreme (e.g. solstice) illumination geometries and overlaps existing spacecraft observations of Earth. Section 6 includes a simulation of earthshine radiance and its use as a natural illumination source for measurements in lunar polar shadowed terrain.

## 2. Earth Spectral Radiance

Earth's radiance is simulated using spectral image cubes generated by the VPL's 3D Earth radiance modeling capability, as described by *Robinson et al.* (*2011*). This is an improved version of code originally developed to simulate the spectra of Earth-like exoplanets (*Tinetti et al., 2006a, 2006b*). The atmospheric portion incorporates a set of gas transmittances using line-lists from the 2008, high-resolution transmission molecular absorption (HITRAN) database (*Rothman et al., 2009*), which includes the major terrestrial absorbing species. Atmospheric temperature profiles and mixing ratios are constrained to season, local time and location using data from multiple Earth-sensing missions, including the Microwave Limb Sounder (MLS) (*Waters et al., 2006*), Tropospheric Emission Spectrometer (TES) (*Beer et al., 2001*) and Atmospheric Infrared Sounder (AIRS) (*Aumann, 2003*). The modeled surface consists of a suite of surface types (forest, grassland, desert, snow, and ocean) whose distributions are derived from multi-year measurements by the Moderate Resolution Imaging Spectroradiometer (MODIS) (*Salomonson et al., 1989; Riggs et al. 1999*). Location and extent of snow cover and sea ice are based on weekly averages derived from MODIS observations. Cloud type, thickness and altitude represent large and unpredictable variability in both Earth solar reflectance and thermal emissivity. The spatial distribution of cloud cover at the date and time of each model are approximated using concurrent MODIS measurements of cloud top properties, or (when no data is available) by interpolation between measurements made temporally close to the model date and time. Ocean reflectance properties including glint are calculated with the Cox and Munk (*1954*) formalism. All of these model elements are stored in a time-dependent atmospheric and surface state library, and this library is then used by the Spectral Mapping Atmospheric and Radiative Transfer (SMART) model (*Meadows and Crisp, 1996; Crisp, 1997*) to generate and map resolved, outgoing spectral radiances extending from far-UV (0.1 μm) to far-IR (200 μm) wavelengths onto a pixelated sphere representing a spatially-resolved Earth.

2.1 Solar-Band Wavelengths

The accuracy of the VPL modeling capability has been validated from observations of Earth obtained in 2008 by the Extrasolar Planet Observation and Deep Impact Extended Investigation (EPOXI) mission – a Discovery-class mission using the repurposed Deep Impact spacecraft instruments (*Livengood et al.,* 2011; *Robinson et al.,* 2011*; Schwieterman et al., 2015*). Earth's radiance was measured over complete rotations from near-UV through IR wavelengths on 18–19 March, 28–29 May and 4–5 June 2008, using the High Resolution Imager Visible Light Camera (HRIVIS) and High Resolution Instrument Infrared (slit) Spectrometer (HRIIR). Observing phase angles (Sun-Earth-observer angles) on these dates were 57.8°, 75.1° and 76.7° respectively. In all three observations, sub-observer latitude was close to the equator (at 1.6°N, 0° and 0°). Whole-disk radiance was measured repeatedly by the camera in seven broad (Δλ ~ 50–100 nm) filter bands covering 365 to 970 nm. The IR slit spectrometer recorded whole-disk spectra from 1100–4540 nm wavelengths by orienting the slit parallel to the equator and repetitively scanning pole-to-pole. Observations were repeated at a cadence of 15 minutes or less for the camera and ~2 hours for the spectrometer, providing longitudinally resolved data sets.

Figure 1 shows a rotation-averaged and disk-averaged VPL model earthshine spectrum (dark trace) that reproduces the 19 March 2008 "EarthObs1" observation (77 % disk illumination). Both sub-observer and sub-solar points lie within a few degrees of the equator. EPOXI measurements (blue and red) are superimposed on the model. Narrow spectral features apparent at shorter wavelengths are mostly the solar Fraunhofer spectrum, with additional telluric absorption features from $H_2O$, $O_3$ and $O_2$ in Earth's atmosphere, particularly the prominent A-band of $O_2$ centered at 0.764 μm. Broad, deep features in the spectrum at long wavelengths indicated by asterisks are water vapor absorption bands. The abrupt cutoff in Earth radiance at short wavelengths ($λ$ < 0.33 μm) is the product of absorption in the very strong 0.255 μm Hartley band of ozone (*Chamberlain and Hunten, 1987*) and rapidly declining solar radiance, partially opposed by a contribution from Rayleigh scattering. The weak spectral structure near 0.6 μm and the rise in reflectance ~0.7 μm is the spectral signature of the vegetation "red edge" discussed below. The green trace shows the reflected sunlight from a Lambert surface for the same solar illumination geometry, where we have used a solar spectral model with approximately the same spectral sampling. A Lambert albedo of ~0.29 provides about the same spectrally integrated irradiance at this phase angle, although the energy distribution is different as a consequence of atmospheric absorption and scattering. This value is close to the accepted value for Earth's planetary albedo (*Stephens et al., 2015*) and within well-defined bounds for Earth's apparent albedo, derived from multi-year, ground-based observations of earthshine at visible wavelengths (*Qiu et al., 2003; Pallé et al., 2003*). A portion of the near-IR spectrum is shown at expanded scale (factor of 10) in order to highlight the good agreement between the VPL model and measurements at longer wavelengths.

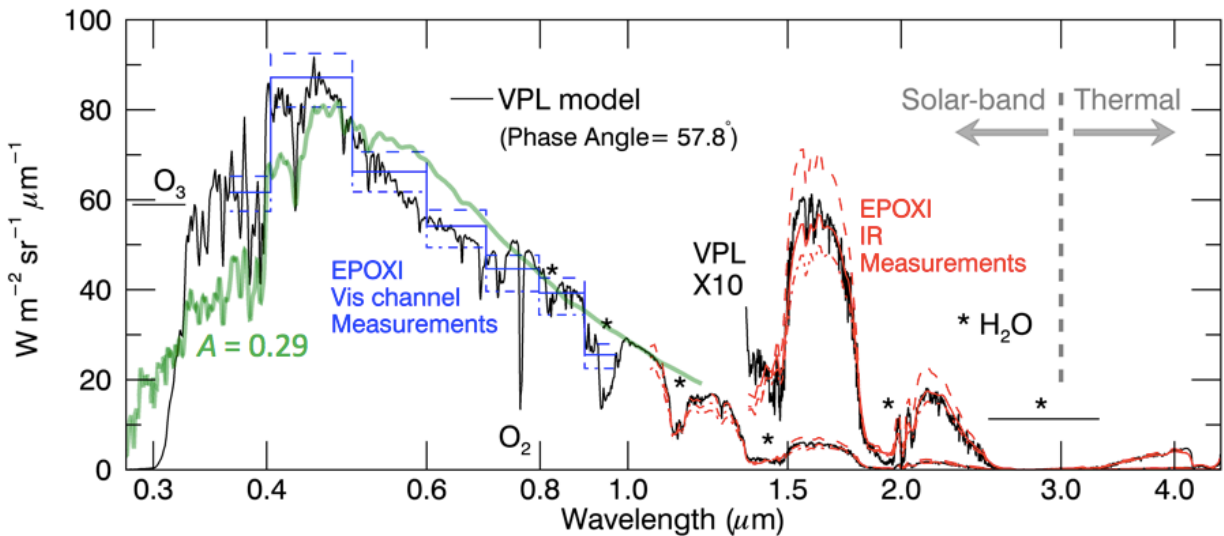

*Figure 1. Disk-averaged spectral radiance of Earth as observed by the EPOXI multi-band camera (blue histogram) and IR spectrometer (red trace) on 19 March 2008. Dashed lines show the extreme values measured over a full Earth rotation. The black line shows the 24-hour average, Virtual Planetary Laboratory (VPL) model spectrum for the same date and observing geometry. Significant water vapor absorption bands are marked with an asterisk (\*). Green line shows the brightness of a Lambert reflector with apparent albedo of 0.29.*

2.2 Thermal Emission Spectrum

A VPL model of Earth's thermal emission spectrum is shown in Fig. 2. Because of its broad spectral extent, the total thermal contribution is a significant part of Earth's overall energy output, even though radiance values are a fraction of that in the solar band. The spectrum is marked by deep and extensive absorption bands from atmospheric $H_2O$, $CO_2$ and $O_3$ that overlie the warm surface. Earth's spatially-dependent spectrum is controlled by the distribution of surface types and the presence of overlying clouds. The peak radiance and the depth of the absorption bands are sensitive to both cloud cover and surface temperature, as demonstrated by the Moderate Resolution Atmospheric Transmission (MODTRAN) models (Climatemodels.UChicago.edu/modtran/) that show cloud-free upwelling spectral radiance at mid-latitudes during summer (red trace) and winter (blue trace). In the absence of clouds, which include a cold cirrus component, the radiance at 11 µm would be ~10% larger. Excluding the deep absorption bands, this sample Earth spectrum near equinox is well matched by a 279 K blackbody, after applying a limb-darkening correction (*Hearty et al., 2009*). Despite a seasonal modulation in one latitude band, it is shown in Sec. 4 that at any one season, the disk-average thermal spectrum is very uniform over a single rotation, as a result of spatial averaging of warm and cold surfaces.

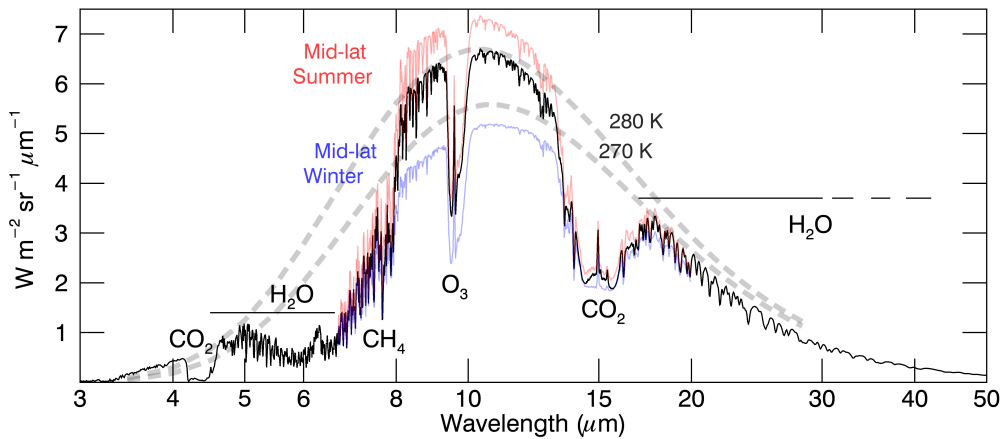

*Figure 2. VPL model spectrum showing Earth's thermal radiance for the observing circumstances of Fig. 1. Red and blue traces show regional upwelling radiances computed using MODTRAN. Included for comparison are representative blackbody radiance curves from a unit emissivity surface at two temperatures.*

To estimate the overall accuracy of the VPL models, it is important to distinguish between: (i) uncertainties in the measurements that validate them, which are estimated to be ~10% in absolute radiance based on calibration uncertainties from EPOXI; and (ii) true variability in Earth's brightness due to clouds, temperature, and the seasonal distribution of ice and snow cover. The VPL 3D spectral Earth models more accurately reproduce relative variations than absolute radiances (*Robinson et al., 2011*). Consequently, variability due to (ii) is well simulated, but unknown radiance offsets may be present as a result of (i).

### 3. Earth Radiance Model Suites

Several suites of VPL Earth models are used in this analysis, covering two full lunations around the 2008 spring equinox:

(1) A set of 3D spectral image cubes of Earth as viewed from the Moon, computed at one-day cadence and spanning a 16-day portion of the March 2008 lunation. This date range brackets the March 2008 EPOXI observation, and coincides with northern spring, when solar insolation is nearly symmetric about Earth's equator. Table 1 lists the observing circumstances on these dates, compiled using the JPL Horizons online ephemeris (http://ssd.jpl.nasa.gov/?ephemerides). Coordinate definitions are shown diagrammatically in Fig. 3a.

(2) Two additional sets of image cubes at 3-hour cadence, each covering one Earth rotation. The first (8 March 2008) is near full-illumination with orbital phase angle $g < 17.2°$, and the second (18 March 2008) is at crescent phase, where $g = 135–147.2°$. Dates are highlighted in gray in Table 1, and the time coverage is indicated by orange shading in Fig. 3b.

(3) A high time-resolution (1-hour cadence) VPL series listing disk-averaged spectral radiance of Earth, from 21 March and 20 April 2008, as Earth's orbital phase angle goes from ~180°–0°–180°.

(4) A series of full-disk VPL spectra from the EPOXI perspective is used to simulate the 19 March 2008 "EarthObs1" observation.

**Table 1.** *Earth observing circumstances as seen from the lunar south pole during March 2008, from JPL Horizons Ephemeris (http://ssd.jpl.nasa.gov/horizons.cgi). Entries correspond to VPL 3D models. Quantities $r_\odot$ and $\Omega_\oplus$ are Earth-Sun distance and Earth solid angle viewed at the Moon. The quantity g is Earth's phase angle, where fractional illumination is defined as $1/2 (1 + \cos g)$. Quantities CML and $\theta$ are terrestrial central meridian W. longitude and latitude of the sub-observer (sub-sensor) point, and $L_\odot$ and $\theta_\odot$ refer to the terrestrial sub-solar point. Shaded entries indicate dates near full phase and crescent phase when VPL models provide diurnal coverage.*

| March day (0 hr UT) | $r_\odot$ [AU] | $\Omega_\oplus$ [×10⁴ sr] | $g$ [°] | Illum. [%] | CML [°] | $\theta$ [°] | $L_\odot$ [°] | $\theta_\odot$ [°] |
|---|---|---|---|---|---|---|---|---|
| 05 | 0.99189 | 8.60 | 34.3 | 91.3 | 209.6 | -19.5 | 177.1 | -6.0 |
| 06 | 0.99215 | 8.83 | 22.0 | 96.4 | 197.7 | -14.2 | 177.2 | -5.6 |
| 07 | 0.99241 | 9.06 | 9.3 | 99.3 | 186.1 | -8.2 | 177.2 | -5.2 |
| 08 | 0.99267 | 9.25 | 4.1 | 99.9 | 174.6 | -1.7 | 177.3 | -4.9 |
| 09 | 0.99293 | 9.39 | 17.2 | 97.8 | 163.0 | 5.0 | 177.4 | -4.5 |
| 10 | 0.99319 | 9.48 | 30.6 | 93.0 | 150.9 | 11.6 | 177.4 | -4.1 |
| 11 | 0.99345 | 9.51 | 44.1 | 85.9 | 138.2 | 17.5 | 177.5 | -3.7 |
| 12 | 0.99372 | 9.47 | 57.6 | 76.8 | 124.7 | 22.4 | 177.5 | -3.3 |
| 13 | 0.99398 | 9.40 | 70.9 | 66.3 | 110.4 | 25.9 | 177.6 | -2.9 |
| 14 | 0.99424 | 9.29 | 84.1 | 55.1 | 95.6 | 27.7 | 177.7 | -2.5 |
| 15 | 0.99451 | 9.17 | 97.2 | 43.8 | 80.8 | 27.6 | 177.8 | -2.1 |
| 16 | 0.99478 | 9.02 | 110.0 | 32.9 | 66.4 | 25.9 | 177.8 | -1.7 |
| 17 | 0.99505 | 8.88 | 122.6 | 23.1 | 52.9 | 22.6 | 177.9 | -1.3 |
| 18 | 0.99532 | 8.73 | 135.0 | 14.6 | 40.3 | 18.1 | 178.0 | -0.9 |
| 19 | 0.99559 | 8.58 | 147.2 | 8.0 | 28.7 | 12.8 | 178.0 | -0.5 |
| 20 | 0.99587 | 8.43 | 159.2 | 3.3 | 17.8 | 7.1 | 178.1 | -0.1 |

The VPL Earth models were created at 1 cm$^{-1}$ spectral resolution, and then reduced in size for this study by coadding into 1167 solar band spectral channels of 40 cm$^{-1}$ width ($\lambda$=0.20–3.00 μm) and 1039 thermal band channels of 3 cm$^{-1}$ width ($\lambda$ = 3.02–49.8 μm). In this study, spectra are displayed on a wavelength scale, although spectral sampling is in equal wavenumber steps since that is better suited for representing prominent spectral features with the fewest number of spectral points. Models list the intensities $I_p$ (W m$^{-2}$ sr$^{-1}$ μm$^{-1}$) from each Earth pixel, computed along the direction to the lunar observer.

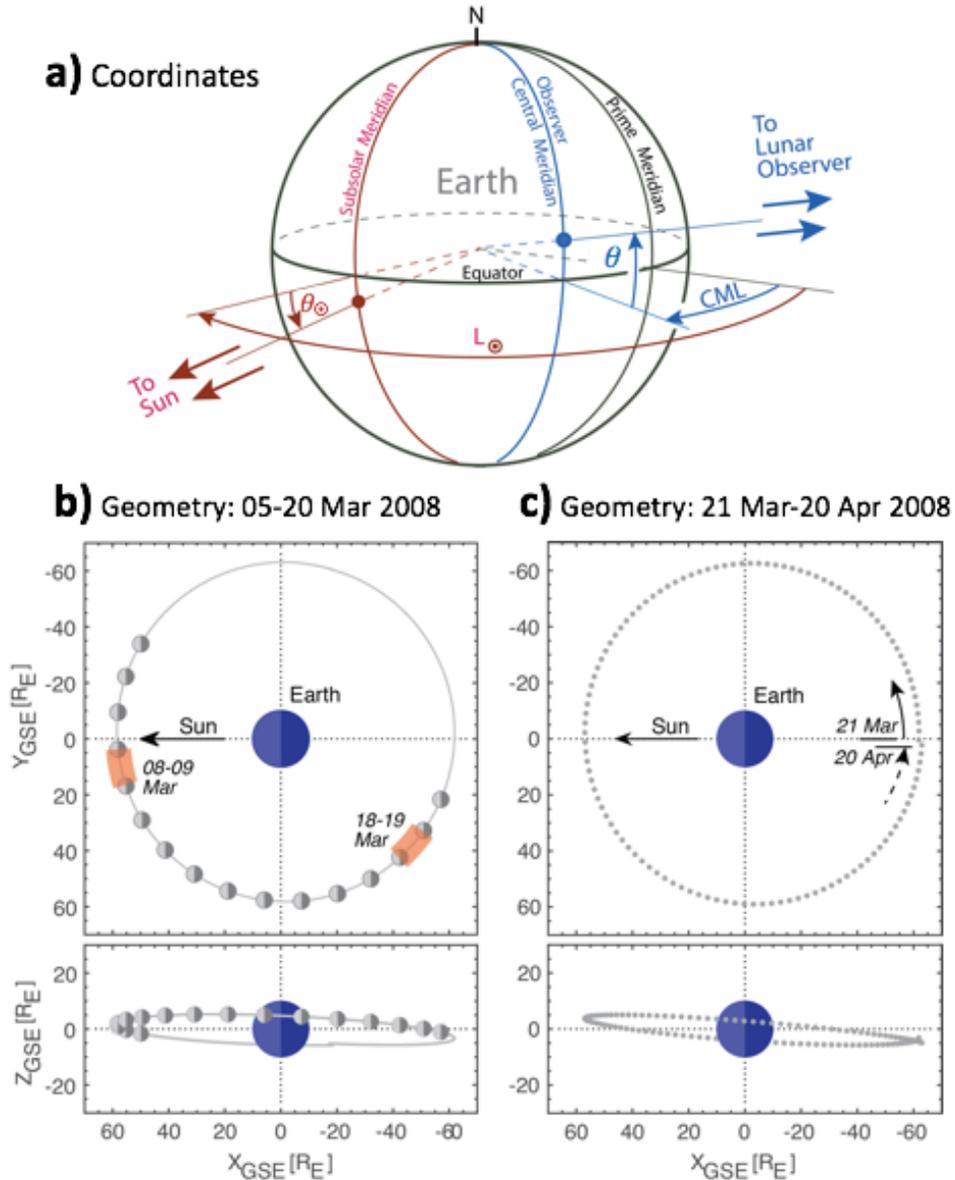

*Figure 3. Earth illumination geometry as observed from the Moon during spring 2008. (a) Diagram illustrating the coordinate geometry in Table 1. (b) Earth-Moon relative positions spanning the full date range of the VPL 3D models. Plots are displayed in Geocentric Solar Ecliptic coordinates (GSE), with X-axis pointing to the Sun and Z-axis pointing normal to the ecliptic plane. Earth and Moon diameters are not to scale. The shaded sectors indicate single Earth rotations modeled at 3-hour time steps near full-phase and crescent phase (shaded entries in Table 1). (c) Relative positions of Earth and Moon spanning the March–April 2008 orbital phase curve at 1-hour cadence. Positions are separated by 6-hour time gaps for clarity.*

### 3.1 Image-Cube Format

Earth's surface is divided into equal area pixels using the HEALpix method for discrete representation of functions over the surface of a sphere (http://healpix.jpl.nasa.gov) (*Górski et al., 2005*). This tessellation method is used here since it is computationally highly efficient and has been adopted as the preferred spatial format in the VPL Earth models. Each of the $N_p$ surface elements (our choice for spatial resolution yields $N_p$ = 768 over the entire sphere) is fixed in latitude and longitude, and assigned a category of surface reflectance. Additional, lower resolution HEALpix layers are superimposed, to account for atmospheric thermal properties, clouds at multiple altitudes, and the spatially-dependent mixing ratios of absorbing gases.

Disk-averaged VPL model radiance $I_\lambda$ in the observer direction is defined by

$$I_\lambda = (4/N_p) \sum_{p_{vis}} \mu_p \; I_p(\lambda) \qquad (1)$$

where $I_p(\lambda)$ *is* the outgoing spectral radiance from each pixel, computed along the sightline to the observer, $\mu_p$ is the cosine of the local zenith angle in the observer direction, and the summation is over observable pixels ($\mu_p > 0$). Other quantities which follow directly from the array elements are full-disk spectral irradiance,

$$J_\lambda = I_\lambda \; \Omega_E \qquad (2)$$

with $\Omega_E$ Earth's apparent solid angle. Solar (or thermal) band-integrated irradiance ***J*** is obtained by summing over the spectral axis,

$$J = \sum_\lambda J_\lambda \; \Delta\lambda_\lambda \qquad (3)$$

The maps in Fig. 4 show the general features of Earth's radiance from near full to crescent phase; at $\lambda$ = 1.0 μm where the atmosphere is mostly transparent, and at $\lambda$ = 11 μm where Earth's thermal emission is largest. Artificially colored and cloud-free "Earthview" maps at the same times and observer perspectives (*www.fourmilab.ch/earthview/*) are included in the top row of the figure for geographic context and to show Earth's illumination phase at the Moon. The short wavelength maps reveal the high reflectance of the major continents and wide-area cloud cover, derived from MODIS measurements at the date and time of the simulations. Earth's thermal flux is nearly independent of phase as discussed below, so it becomes the main radiance contributor for a portion of each lunation. Most of the thermal radiance is confined to warmer latitudes. This latitude gradient will need to be taken into account in shadowing codes that attempt to estimate the thermal irradiance from Earth when it is rising or setting, especially near the lunar poles, where the gradient is perpendicular to the local horizon.

### 3.2 Distinctive Surface Spectra

Spectral endmembers belonging to dominant surface types show up in the VPL image cubes, as illustrated in Fig. 5. These can be identified by combining information from reflectance, thermal and artificially colored context images. The example in Fig. 5 shows the reflectance spectrum of South American forest land (panel e), which is marked by strong absorption from vegetation up to $\lambda \sim 0.7$ μm, followed by an abrupt increase in reflectance at longer wavelengths, the so-called "red edge" of plant chlorophyll (*Filella and Panuelas, 1994; Tinetti et al., 2006a; Livengood et al., 2011*). Other contributors are high and low altitude clouds (panels f and g), which can be distinguished based on temperature and the depth of $H_2O$ bands (lower clouds have more overlying water vapor), and ocean surface (panel h), which, except for Sun glint at high phase angle, shows low reflectance at visible and near-IR wavelengths. Other distinct surface types (e.g., desert, ice, snow cover and grasslands) are included as components of the VPL model. Although not conspicuous here, they are present and would show up prominently in VPL models at other longitudes or seasons.

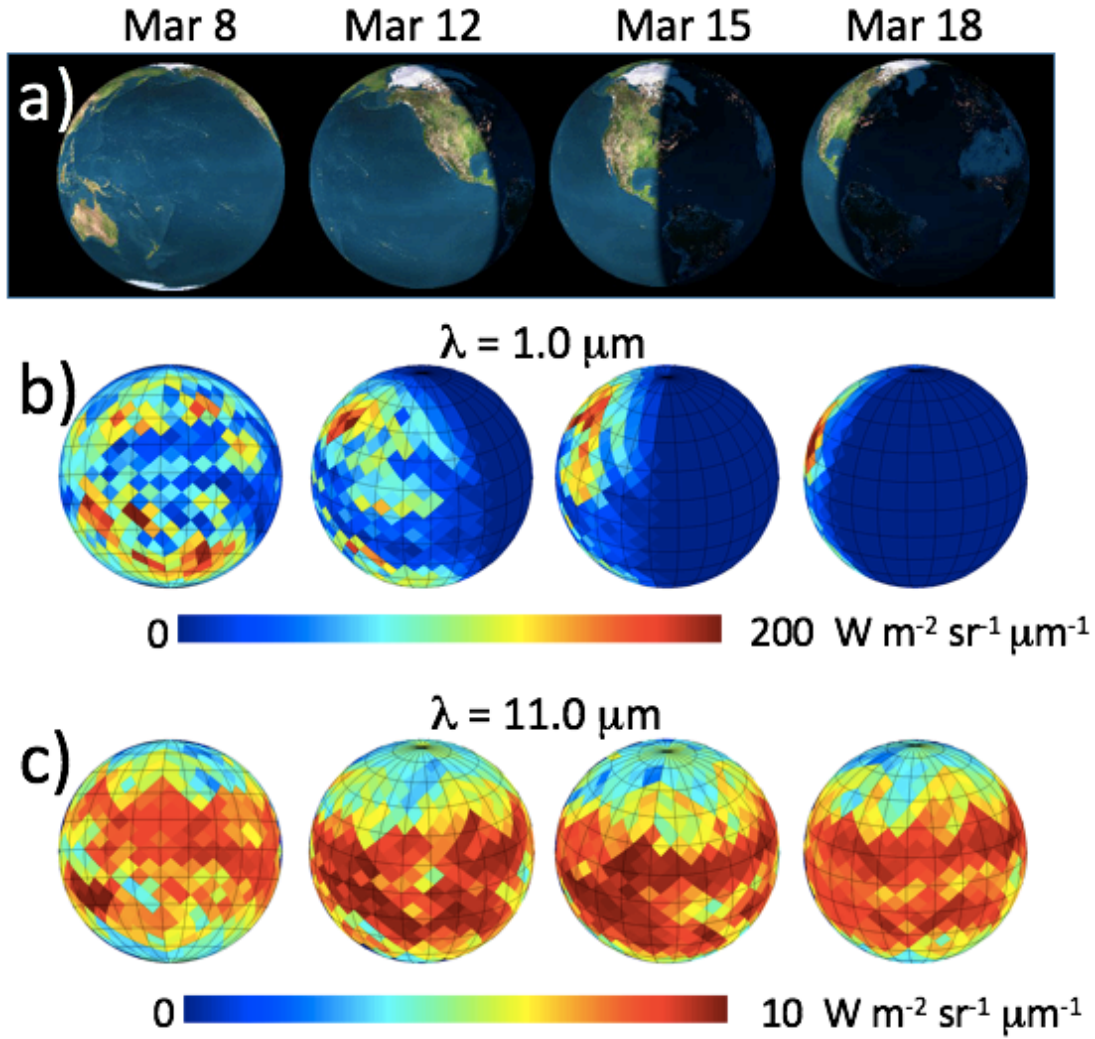

*Figure 4. Spatial distribution of Earth's radiance observed from the Moon during four days in March 2008 at 0 HR UT. (a) Artificially colored and cloud-free context maps (http://fourmilab.ch/earthview) show solar phase and terrestrial sub-observer location. (b) λ= 1.0 µm brightness maps show the brightness of surface features; continents, cloudy regions and ocean glint near crescent phase. (c) Thermal maps reveal the strong contribution from higher land and ocean temperatures at low and mid-latitudes.*

## **4. Diurnal and Orbital Light Curves**

Earth's broadband, hemispherical irradiance at the Moon was computed daily at 0 HR UT over the duration of the March 2008 lunation, as shown in Fig. 6. Solar-band flux (blue dashed lines) naturally depends on Earth's phase, and declines from the peak value of 60–75 mW m$^{-2}$, depending on facing longitude, near zero phase (8 March 2008) to a small value on 20 March. Ironically, the daily 0 hr UT sequence missed those longitudes with larger fractional landmass, as shown in the inset images, coinciding with a minimum in disk-averaged reflectance. The region between vertical dotted lines shows the longitude dependence of brightness during the 8–9 March rotation with 3-hour time step. Peak brightness occurs as the observer's CML traverses the African continent and portions of Asia between ~300°W and ~0°W.

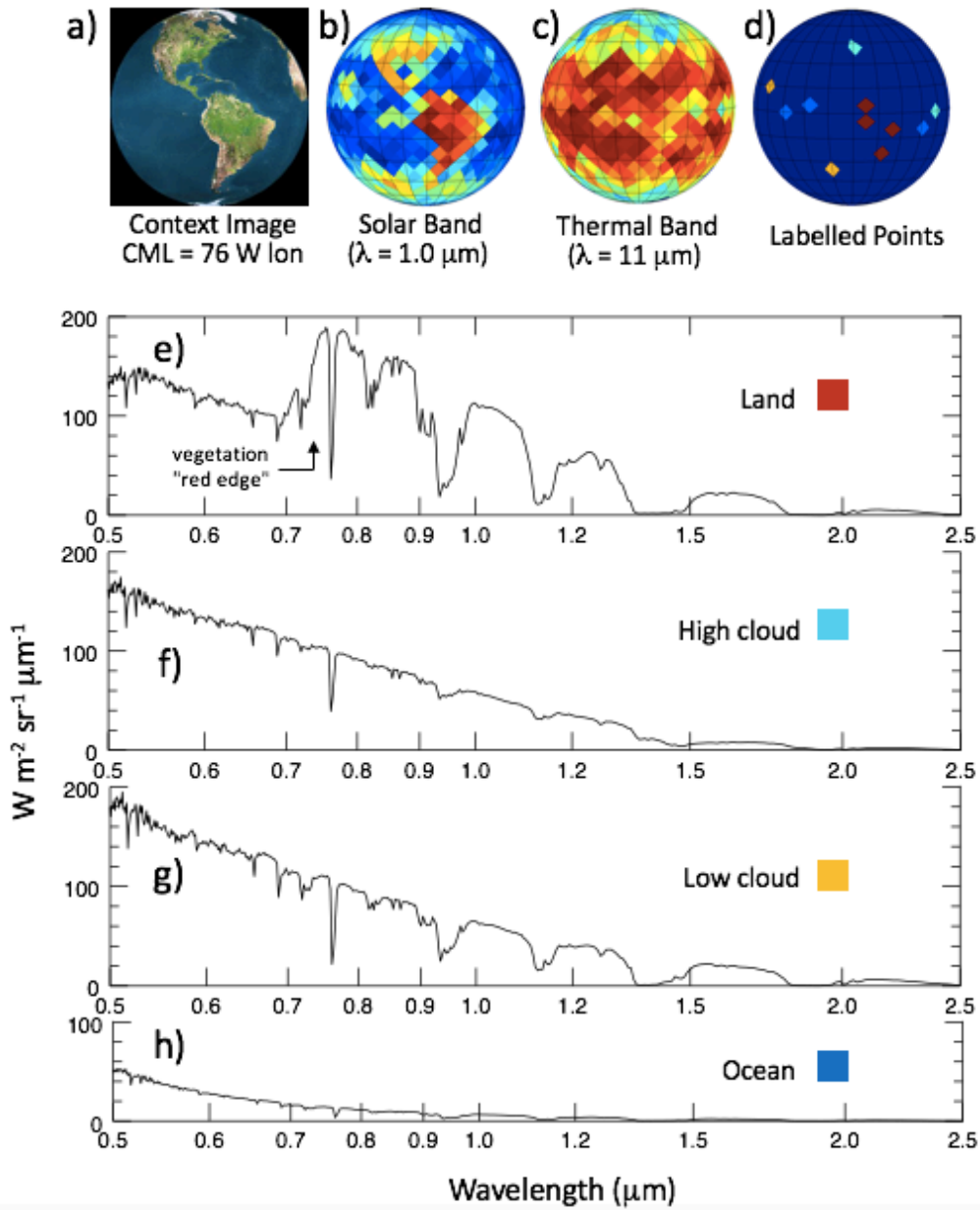

*Figure 5. Representative point spectra of distinctive surface types. (a) Context image showing continent outlines at sub-observer longitude of 76°W (CML = 76°). This, along with HEALpix maps showing (b) reflectance at 1 μm and (c) thermal emission at 11 μm are used together to distinguish between surface types. These include: (e) land, which exhibits high reflectance, the distinct "red edge" of plant chlorophyll near 0.7 μm, and high thermal radiance; (f) high clouds, with strong reflectance blue slope and low thermal radiance; (g) low clouds, with deep $H_2O$ absorption bands and higher thermal radiance; and (h) ocean surface, with low reflectance and high thermal radiance. The locations of these points are shown color coded in panel (d).*

Whole-disk thermal emission (60–70 mW m$^{-2}$ over the time span of Table 1) shows very little change with observer longitude, since variations in surface emissivity between land and ocean are much smaller than variations in solar-band reflectance. A portion of the observed trend in irradiance visible in Fig. 6 is a consequence of Earth's changing angular size, which is largest on 11 March. Cold cirrus clouds provide the largest thermal contrast with the warmer surface, but these appear sporadically and typically cover ~20% of the surface or less. Such small changes in the disk-averaged radiance are perhaps surprising, since sub-observer latitude moves from 20°S to 28°N over the model date range. Despite the wide spread in thermal radiance between low and high terrestrial latitudes (Fig. 4c), the relative contributions from warm and cold regions do not change significantly at this season.

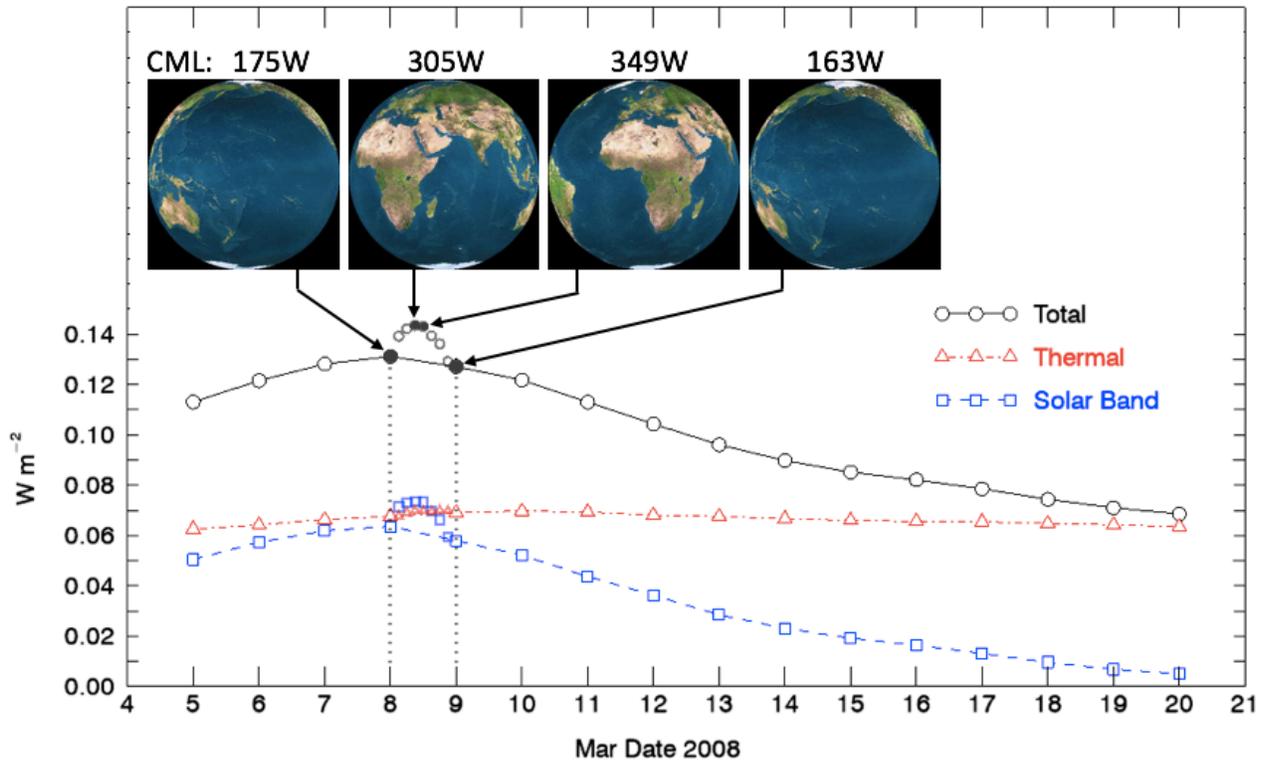

*Figure 6. Brightness curves showing Earth's broadband, hemispherical irradiance at the Moon during the March 2008 lunation. Dashed blue line shows solar band reflectance, which declines from a maximum near zero phase angle on 8 March to a small value at crescent phase. Thermal emission (dashed red line) is nearly independent of sub-observer longitude (CML). The 8 March sequence (region between dotted lines) shows longitude sampling at 3-hour cadence near zero phase. The band of longitudes between ~300°W and 0°W, which were missed during the daily 0 HR UT sequence, show a reflectance maximum that is mostly a consequence of the larger visible landmass fraction.*

4.1 Spectral Dependence

The spectral distributions of the diurnal brightness extremes are illustrated in Fig. 7. The greatest contrast between land and ocean surface occurs at near-IR wavelengths beyond ~0.7 μm, and in the regions between strong water vapor absorption bands that obscure the surface and lower atmosphere. Virtually no brightness contrast appears at wavelengths < 0.45 μm as a result of extinction in the overlying atmosphere from a combination of strong ozone absorption and Rayleigh scattering.

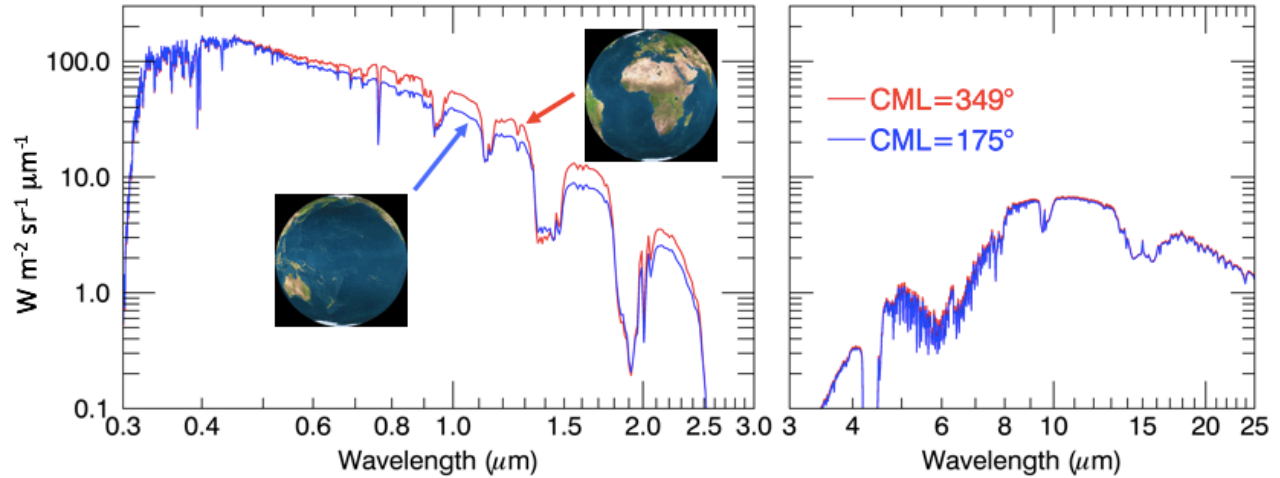

*Figure 7. Diurnal extremes in hemispherically averaged radiance from Earth's near fully (> 99%) illuminated disk. The red spectrum shows bright reflectance resulting from a larger fraction of visible land coverage, whereas the blue trace shows the spectrum when the facing hemisphere is largely ocean-covered. Cloud reflectance, not shown in the artificial images, contributes to both cases. Most of the brightness difference occurs at near-IR wavelengths, and between atmospheric $H_2O$ absorption bands, where the atmosphere is transparent. Reflectance at near-UV wavelengths is dominated by Rayleigh scattering and $O_3$ absorption, which largely obscures the surface.*

Figure 8 shows Earth's complete brightness curves at the Moon at discrete wavelengths as a function of phase angle, derived from the one-hour grid of hemispherically averaged intensities between 19 March and 22 April 2008. Dashed lines show daily average values over one Earth rotation, convolved using a 24-hour sliding "boxcar" averaging window, which we designate $\overline{I_\lambda}$. These wavelengths were selected to highlight the transition from short wavelength scattering plus $O_3$ absorption (panel a), to a mostly clear atmosphere at near-IR wavelengths and away from water vapor absorption bands (panels c and d). The central parts of the light curves in panels (c) and (d) show the strong diurnal brightness modulation, which peaks as Earth's largest continents rotate through the observer's central meridian. Weaker coherent fluctuations can also be observed at 0.4 µm, in addition to random variability from clouds.

Especially interesting is the appearance of the light curve at longer wavelengths. Significant reflectance persists during Earth's crescent phases, as a consequence of both ocean glint (*Cox and Munk, 1954; Williams and Gaidos, 2008; Robinson et al., 2010, 2014*) and cloud forward scattering. Ocean glint produced by specular reflectance of sunlight on the ocean surface contributes to disk brightness at all phases, but its relative importance increases dramatically at crescent phase when most of the disk is dark. This effect is also illustrated in Fig. 9, which shows a sequence of model images spanning an Earth rotation at crescent phase. The glint effect remains important to phase angles approaching ~160°, at which point the distribution of slopes on the ocean surface suppress the glancing reflection of sunlight. The contribution of clouds to the overall brightness and its variability increase when observed at large phase angles. The results of scattering computations show that aerosols of a few µm radius, typical of cloud water droplets, strongly scatter light into the forward direction and begin to increase in brightness when phase angles exceed ~120° (*Hovenier and Hage, 1989*). This effect becomes pronounced at longer wavelengths since (for a fixed aerosol size distribution) the angular width of the single-scattering phase function increases, i.e., cloud forward scattering is observed at smaller phase angles.

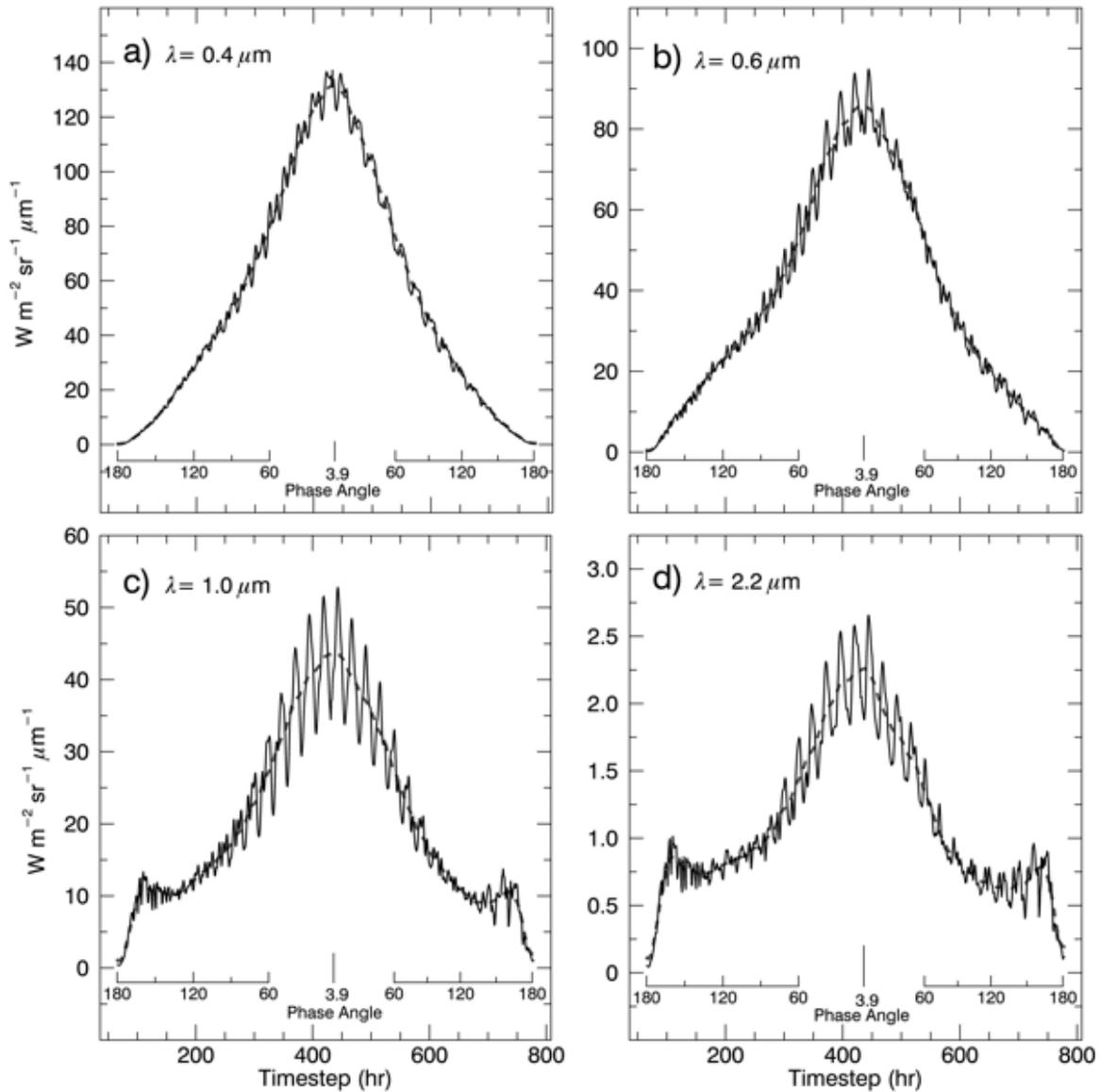

*Figure 8. Earth's full light curve at multiple wavelengths, from the March–April 2008 VPL time series, computed at 1-hour cadence. As in Fig. 7, the vertical scale is hemispherically averaged radiance. Dashed lines show the light curves averaged over one Earth rotation. (a) Light curve at 0.4 μm. Optical depth of Rayleigh scattering is sufficient to block most of the surface light. (b) At longer wavelengths, optical depth diminishes and the surface becomes partially visible. (c) The phase curve is dominated by the diurnal modulation of surface reflectance. Ocean glint and cloud scattering become important at crescent phase. (d) The modulation of surface reflectance by changing land and cloud coverage is still significant. Cloud forward scattering dominates at large phase angles and is the main source of unpredictable reflectance variability.*

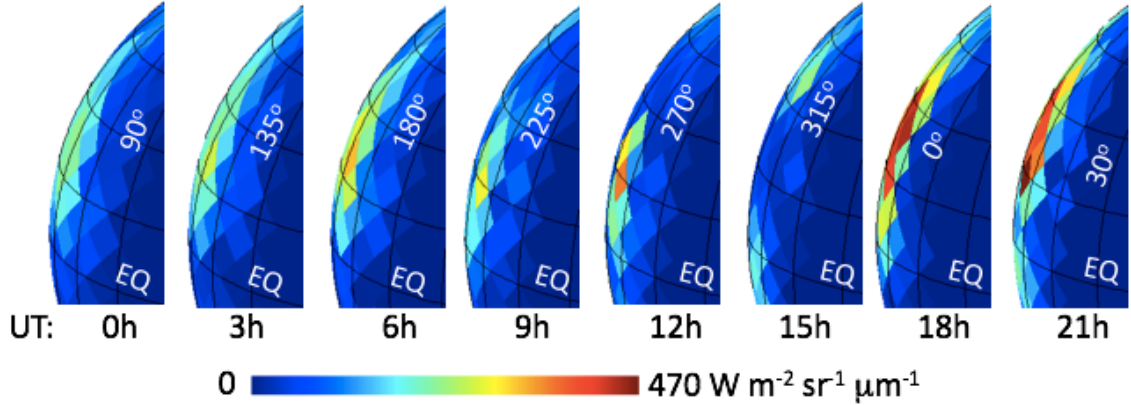

*Figure 9. VPL model image sequence (λ = 1.0 µm) spanning one Earth rotation that shows ocean glint and clouds near the limb during Earth's crescent phase on 18 March 2008. Phase angles range from 135° (left image) to 146° (right image). Latitude/longitude lines are separated by 15°. For an ocean-covered limb, models show a systematic increase in peak glint brightness over this range of phase angles. Additional bright clouds are present at northern latitudes in the 18 and 21 hr UT images. Glint in the 15 hr image is obscured by the African continent.*

## **5. A Seasonal Approximation for Solar-Band Earthshine**

Because Earth's solar-band radiance is highly variable, quantitative predictions with better than order-of-magnitude accuracy require the use of radiance models. Despite the overall accuracy of the VPL models, the 3D source computations are elaborate and require supplementary satellite measurements coincident with the model date and time. It is useful to have a semi-empirical approximation for the spectral irradiance of an unocculted Earth that can be obtained using only ephemeris information, valid for phase angles which are "not too large". At near-UV and short visible wavelengths, Earth's light curve is dominated by scattering and can be estimated using a suitable scattering function integrated over the illuminated disk. At longer wavelengths, the diurnal modulation due to changing fraction of visible land surface, as well as short-lived trends from clouds, is superimposed on this light curve, as in Fig. 8.

To construct this approximation, we use the portion of the March–April 2008 VPL time series between waxing and waning gibbous phases, where $g < 60°$, and we assume that brightness change is proportional to change in observable land fraction. Clouds also show a time-variable signature in this series, but because they are transient, they must be treated as random and a source of uncertainty in our approximation. At each of the $N$ time steps $i$ ($N_i = 215$) the VPL reference model $V_{i,\lambda}$ provides full-disk radiance as a function of wavelength. Each model includes a file header that lists visible fractions of major surface types, namely land (forest, grassland, ground and snow/ice), clouds (thin and thick water and ice clouds) and ocean. Figure 10 shows the fraction of the visible, illuminated disk occupied by the land categories, resampled onto a 2D map. This picture is incomplete, since sub-observer latitude, defined as $\theta$, tracks northward during the sequence and is itself a function of $g$. The map, therefore, represents a range of small phase angles. Land fraction shows a well-defined maximum between ~270° and 360° CML, as the major continents rotate past the sub-observer point, and this peak shifts in longitude as $\theta$ increases. Figure 10b shows the diurnally averaged land fraction as a function of sub-observer latitude, i.e., the average of the map values collapsed along the horizontal axis.

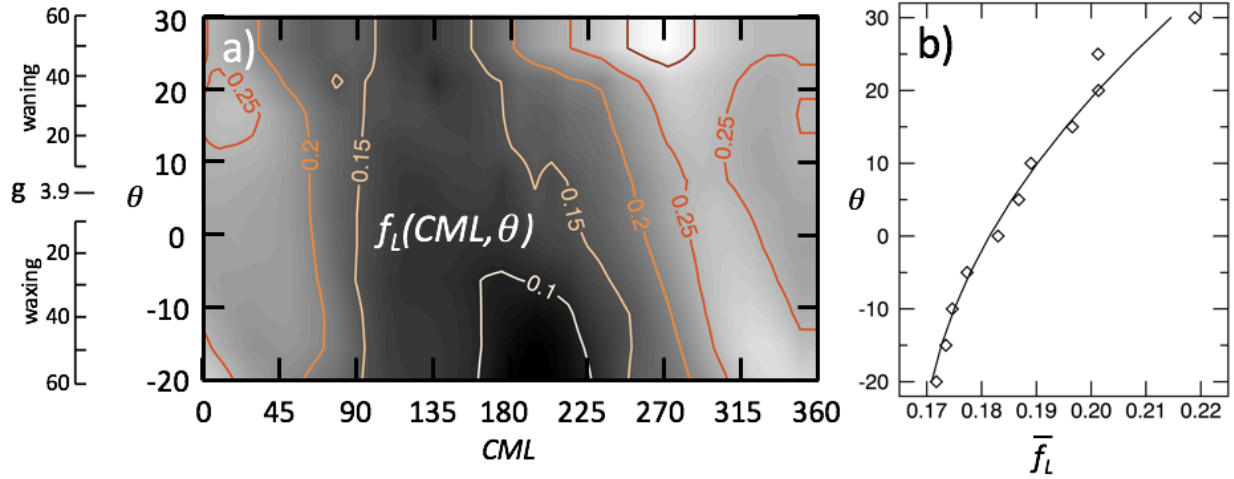

*Figure 10. Fractional land surface $f_L$ on the illuminated visible disk near full phase ( $g < 60°$), from the March–April 2008 sequence of VPL models. (a) Values of $f_L(CML,\theta)$ resampled onto a regular grid of sub-observer latitude $\theta$ and W. longitude. (b) the changing diurnal average value $\bar{f}_L$ as sub observer point moves northward, along with a quadratic fit.*

A similar map analysis (not shown) was carried out for the cloud distribution in order to anticipate how clouds will influence our brightness approximation. Clouds in the VPL series are more distributed over Earth's surface, but they do show broad enhancements that are correlated with ocean surface.

5.1 Diurnal-Average Light Curves

For Earth's rotationally averaged light curve, we assume a fitting function $\bar{I}_\lambda(g,\theta)$, that is the product of the equatorial average radiance at zero-phase, $\bar{I}_\lambda(g=0,\theta=0)$, a light curve shape $\Phi(g,b_\lambda)$, normalized to unity at $g=0$, and a third factor with coefficient $A_{L,\lambda}$ that reproduces the asymmetry in the light curve around $g=0$. This asymmetry results from changing land fraction as sub-observer latitude $\theta$ moves northward during the series. The "—" accent notation is used to represent rotationally (diurnally) averaged quantities.

$$\bar{I}_\lambda(g,\theta) \approx \bar{I}_\lambda(0,0)\ \Phi(g,b_\lambda) \left[1 + A_{L,\lambda}\ \Delta\bar{f}_L(\theta)\right] \qquad (4)$$

In Eqn. 4, we have used abbreviated notation $\Delta\bar{f}_L(\theta) = \bar{f}_L(\theta) - \bar{f}_L(\theta_{g=0})$. The phase integral $\Phi$ is given by

$$\Phi(g,b_\lambda) = (1/2) \prod_{HG}(g,b_\lambda)\ [1 + cos(g)] \qquad (5)$$

The quantity $\prod_{HG}$ is a double-lobed Henyey-Greenstein phase function (*Henyey and Greenstein, 1941; Hapke, 2012, Sec. 6.3*), normalized to unity at $g = 0$. The quantity $b_\lambda$ controls the lobe width ($0 \le b \le 1$). This form is more suitable than a Lambertian phase function because of its ability to adjust to the light curve width. In principle, $\prod_{HG}$ can account for net back-scattering (surface reflectance), or net forward-scattering as with thick haze or in $H_2O$ bands. Numerical testing showed that this shape was generally insensitive to forward-to-back ratio, so equal forward- and back-scattering contributions are used.

At each wavelength, we seek a constrained least-squares solution for $\bar{I}(0,0)$, $b$ and $A_L$ that minimizes the objective function $\sum_i(\bar{V}_i - \bar{I}_i)^2$, with $A_L \ge 0$ and $0 \le b \le 1$. The quantity $\bar{V}_i$ is the running average of $V_i$ over an Earth rotation period. Since the fitting function includes a non-linear factor ($\Phi$), a single-step linear least squares inversion is not applicable. Instead, we use the iterative, Levenberg-Marquardt gradient search algorithm "mpfit" (*Marquardt, 2009*). This accepts user input, including constraints, and returns the set of best-fit coefficients and their uncertainties, along with fit residuals and covariances.

Initial estimates for the coefficients are created by setting $\bar{I}(0,0)$ equal to the maximum value of $\bar{V}_i$, and (using a root-solver) choosing $b$ so that the full-width of the phase integral equals that of $V_i$ near the phase angle limit ($g = 60°$). Using these estimates, $A_L$ can then be solved for via a simple least squares minimization: We rearrange Eqn. 4 to define a new dependent variable $y_i$,

$$y_i = \frac{\bar{V}_i}{\bar{I}(0,0)\,\Phi_i(b)} - 1 = A_L\,\Delta\bar{f}_{L,i} \tag{6}$$

resulting in $A_L = \sum_i \Delta\bar{f}_{L,i}\, y_i / \sum_i (\Delta\bar{f}_{L,i})^2$. At all wavelengths, the starting estimates were found to lie close to the final solution. The uniqueness of the retrievals was verified by mapping the objective function at discrete wavelengths with $A_L$ fixed at the initial value, in order to confirm that it is free of high-order structure and has a single extremum over the $b$ and $\bar{I}(0,0)$ domains. Some correlation (~0.79) does exist between coefficients $\bar{I}(0,0)$ and $b$, but this does not affect the consistency or quality of the retrievals.

Figure 11 shows the VPL diurnal-average curves at several wavelengths (black lines) and least squares fits of Eqn. 4 to these curves using the "mpfit" algorithm (red dashed lines). At visible and red wavelengths, the curves show a clear brightness asymmetry between waning and waxing phases. This is accounted for by increasing diurnal land fraction as sub-observer latitude moves northward. It becomes obvious when compared to symmetric fit results that have $A_L$ fixed at 0 (blue dashed lines). The asymmetry disappears at short wavelengths due to large scattering optical depth, and also at IR wavelengths where the reflectance of spatially variable clouds increases in importance and obscures the brightness change associated with land. At 2.2 μm, clouds distort the light curve shape to the point where $A_L$ retrievals do not converge and returned values of $A_L = 0$.

The plots in Fig. 12 show the coefficient spectra and their two-sigma ($2\sigma$) uncertainties, with $\sigma$ the standard errors returned by the fitting routine. Two-$\sigma$ error estimates are perhaps more realistic, since that represents the ~95% confidence interval for the coefficient values, given "good" models and normally distributed measurements. The width coefficient $b$ is generally tightly constrained, since phase function width exerts a strong influence on the light curve shape. Within deep $H_2O$ bands, the surface component mostly disappears, and the light curve broadens (smaller $b$) due to cloud forward-scattering. The sharp increase in $b$ at the shortest wavelengths ($\lambda < 0.33$ μm) is real and is a consequence of Rayleigh scattering behavior. Retrieved values for $A_L$ are largest at visible and red wavelengths, where surface reflectance dominates and atmospheric extinction is minimal. At longer wavelengths, brightness change becomes more strongly influenced by clouds. Fig. 12d shows the fractional uncertainty in the light curve approximation. This can be obtained from the individual coefficients and their error estimates, but it is simpler and equally valid to use the standard deviation between the approximations and VPL reference model,

$$\frac{\sigma_{\bar{I}}}{\bar{I}} = \sqrt{\frac{1}{N-3} \sum_i \left(\frac{\bar{V}_i}{\bar{I}_i} - 1\right)^2} \tag{7}$$

with $N - 3$ the number of degrees of freedom. This result is represented in the lower plot of panel d. The upper plot shows the total light curve uncertainty $\Delta\bar{I}/\bar{I}$, which includes the ~10% instrumental measurement error in the VPL models (Sec. 2.2) added in quadrature.

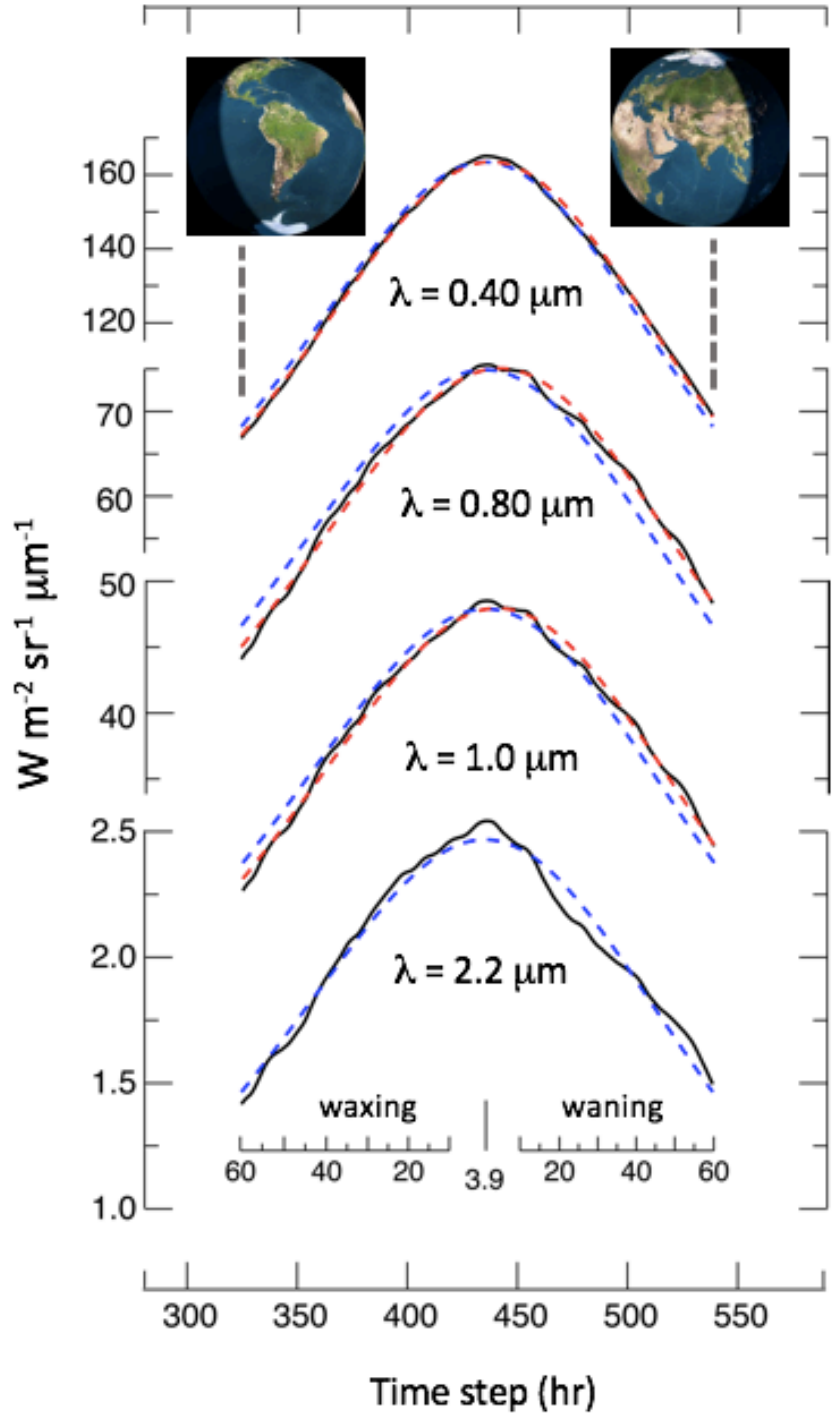

*Figure 11. Longitudinal (diurnal) average VPL brightness curves of Earth from the March–April 2008 sequence (black lines), compared with the light curve approximation (Eqn. 4, red dashed lines). Blue dashed lines are fit results using only the scaled phase function and are therefore constrained to be symmetric about minimum phase angle.*

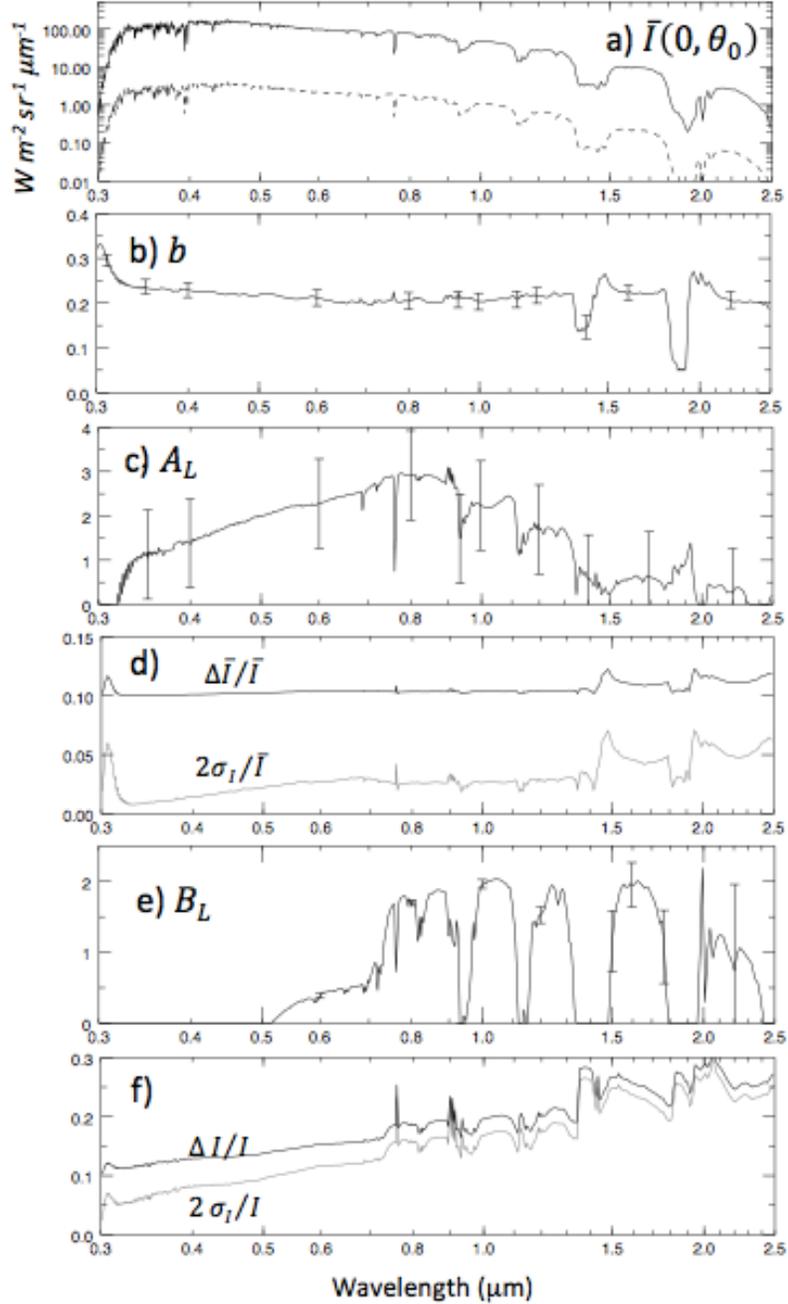

*Figure 12. Coefficients in Eqns. 4 and 8 describing Earth's radiance at phase angles g ≤ 60°, from the April 2008 time series of disk-averaged intensities. Except for panels d and f, uncertainties are twice the standard errors returned by "mpfit". (a) Diurnal-averaged brightness at zero phase angle with dashed lines indicating uncertainty. (b) Double-Henyey-Greenstein broadening coefficient b. (c) Amplitude of the Earth's light curve asymmetry term about zero phase angle. (d) Fractional uncertainty in the diurnal-average light curve; the top trace includes the 10% measurement uncertainty in the VPL 3D source models, added in quadrature. (e) Amplitude of the diurnal modulation term, with shape described by longitude-dependent land fraction. (f) Overall fractional uncertainty in the Earth radiance approximation.*

### 5.2 Diurnal Modulation

To include the diurnal component, we again use projected land area, this time as a proxy for the modulation of radiance about the daily mean,

$$I_\lambda(CML, g, \theta_s) \approx \bar{I}_\lambda(g, \theta_s) \left[1 + B_{L,\lambda} \, \Delta f_L(CML, \theta) \right] \quad (8)$$

with $\Delta f_L(CML, \theta) = f_L(CML, \theta) - \bar{f}_L(\theta)$. At each wavelength, the scale factor $B_L$ and its standard error is determined by "mpfit", with all other coefficients now fixed. Fig. 12f shows the fractional uncertainty in the full brightness approximation at each wavelength, using notation similar to Fig. 12d.

Figure 13 compares the VPL time series with the approximations (red dashed lines) at a few sample wavelengths. Smooth curves show the diurnal average results, as in Fig. 11. Error bars have been omitted for clarity. At wavelengths up to ~1.6 μm, diurnal brightness modulation is well-represented by the observed land fraction. Some additional contribution from clouds is also present. At longer wavelengths and between $H_2O$ bands, the relative brightness of clouds increases. At 2.2 μm, most of the modulation is attributed to longitudinal fluctuations in cloud cover over a range of altitudes. At all wavelengths, discrepancies between the VPL model and the approximations are bracketed by the total uncertainty, which increases to 25–30% beyond 2 μm (Fig. 12f). Tabulated values of all coefficients, useful for estimating Earth's radiance at the Moon, are available as an online data file (Glenar et al., 2018).

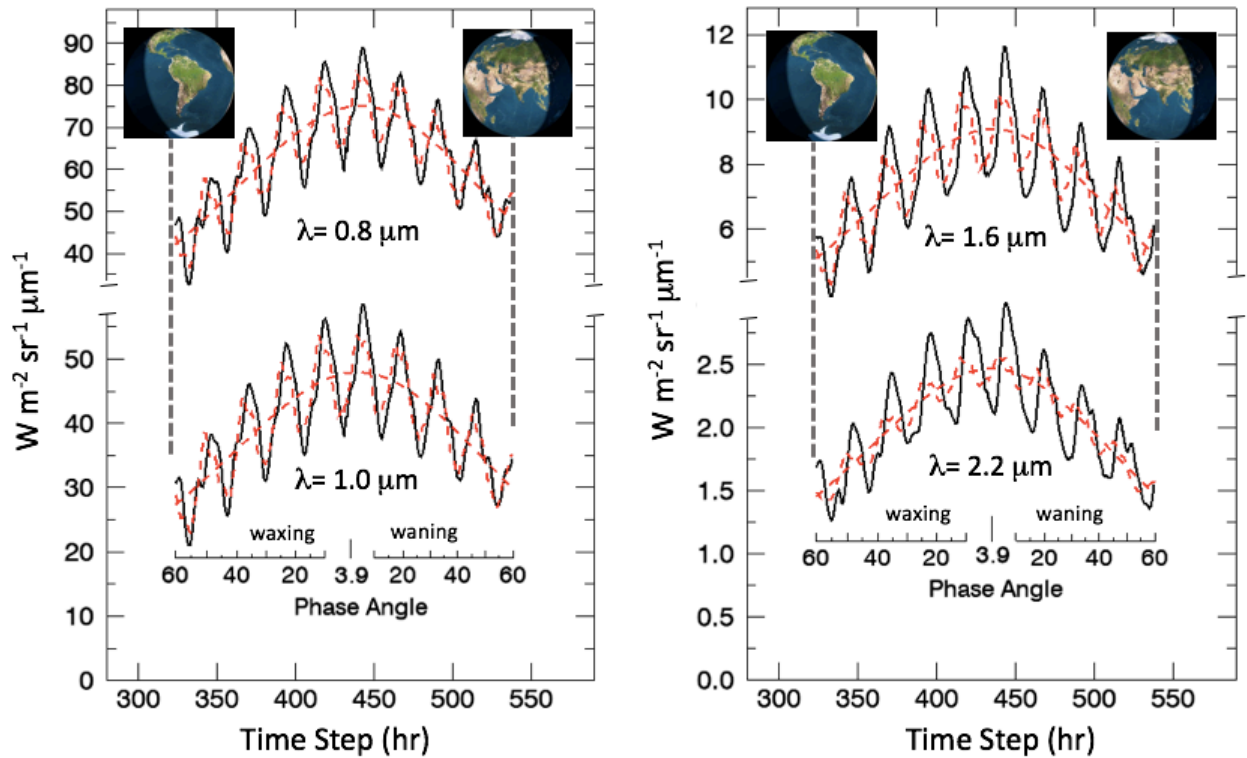

*Figure 13. Earth radiance simulations during the March–April 2008 VPL sequence (black lines), compared with the seasonal approximation (red dashed lines) at four sample wavelengths. Smooth lines show the diurnally averaged phase curves. Clouds show a distinctly different longitudinal distribution than land, and become increasingly bright at long wavelengths. As a consequence, the contribution of the land component to the diurnal brightness modulation diminishes.*

5.3 Comparisons with EPOXI Measurements

We have tested the seasonal approximation by direct comparison with EPOXI measurements, as well as with VPL models that simulate those measurements. During "EarthObs1", EPOXI acquired full diurnal coverage at 57.5–57.8° phase angles and sub-spacecraft latitude of ~1.6°N, within the validity limits of our seasonal approximation. Earth's irradiance was computed from the EPOXI observing perspective and compared against the measurements as shown in Fig. 14. Panels a and b compare measurements from the "VIO" and "NIR" camera channels shown by black circles (*Livengood et al., 2014*), to predictions from Eqn. 8 (red lines), after weighting the computed intensities over the HRIVIS filter bandpass shapes and CCD spectral response. Included for comparison are the VPL simulations, after applying the same spectral weighting steps. Shortward of $\lambda \sim 0.5$ μm (Fig. 14a), there is no diurnal modulation term in our approximation, since retrieved values of $B_\lambda$ are zero.

In all four of these examples, the EPOXI measurements, VPL models and the seasonal approximation are in mutual agreement, to within the overall uncertainties. At near-IR wavelengths (Fig. 14b), both VPL model and approximation predict a larger modulation amplitude than the EPOXI camera measurements. The reason for this disagreement is not clear: It may arise from the present classification of surface types in the VPL Earth model and the spectral albedos assigned to those types. That aspect of the VPL models is currently being investigated. At 2.2 μm (Fig. 14d), diurnal brightness change is controlled more by clouds than by land fraction. As expected, both EPOXI measurements and VPL predictions are in broad agreement, since the VPL source models include the global cloud distribution at the time of the simulation. In our approximation the influence of clouds is reflected in the length of the error bars.

## 6. Discussion

As a natural illumination source, the brightness of visible wavelength earthshine on the lunar surface is generally under-appreciated. For a lunar observer, it would appear some 20–30 times brighter at visible wavelengths than would the Moon at similar phase over a terrestrial landscape, due to both its larger apparent size and higher albedo. This makes earthshine a viable illumination source for quantitative imaging and spectroscopy of the lunar surface in regions either permanently or temporarily devoid of sunlight. One well-documented example is the high signal-to-noise-ratio (SNR) image of the Earth-illuminated lunar limb acquired by the Clementine star tracker in a single, short (0.4 s) exposure (*Hahn et al., 2002, Fig. 4c*). What has been lacking is a calibration of Earth's radiance, which we have created in this study for one season, using a semi-empirical approximation derived from a suite of VPL models.

The broadband irradiance of earthshine also appears to be large enough to influence the stability and transport of some volatiles that may be present on the Moon. Earth's normal irradiance ranges from ~60 mW m$^{-2}$ (thermal contribution alone) to nearly 150 mW m$^{-2}$ near zero phase angle, corresponding to equivalent temperatures of 32–40 K for a unit emissivity surface in thermal equilibrium. During each lunation, equatorward-facing slopes at high latitudes can receive a large fraction of this irradiance, which is many times the average internal heat flow of 9–13 mW m$^{-2}$ estimated by *Siegler and Smrekar* (*2014*). Several volatiles with sublimation temperatures near 20 K ($N_2$, $O_2$, CO, Ar and $CH_4$) are readily mobilized by the thermal contribution from earthshine alone. Other volatiles, including water, have higher sublimation temperatures (Paige et al., 2010) but they may be subtly influenced by earthshine when combined with internal heat flow and scattered sunlight from nearby terrain. The cumulative effect of earthshine should be accounted for in thermal models investigating the long-term stability of volatiles in PSRs. These arguments may even apply to water ice, although no evidence of Earth's influence was observed in a recent survey of Moon Mineralogy Mapper data, which documented ice bearing locations at lunar high latitudes (*Li et el. 2018*).

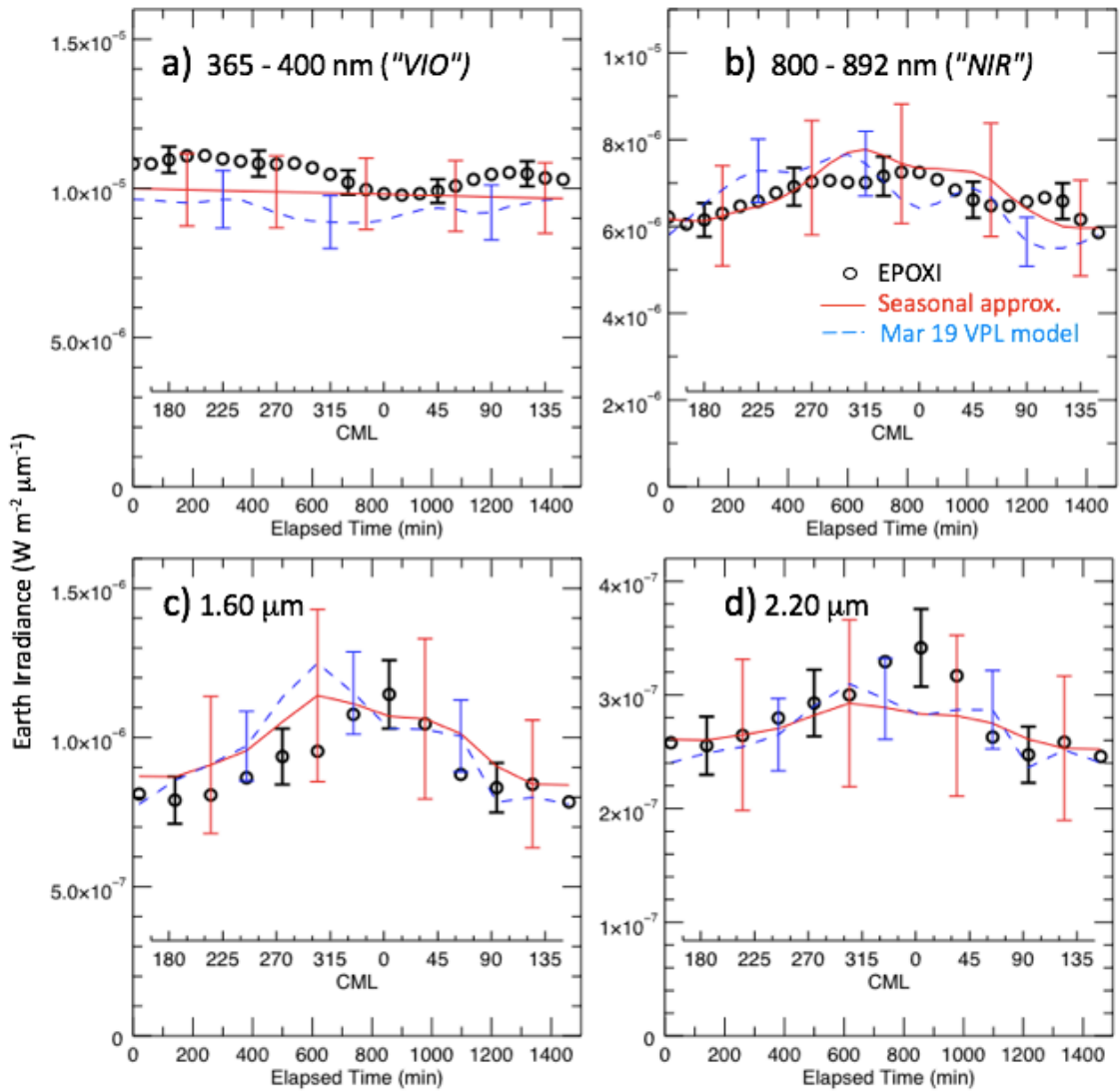

*Figure 14. Comparison of the 19 March 2008 EPOXI "EarthObs1" measurements (g = 57.5–57.8° and Earth-spacecraft distance of 0.182–0.185 AU) covering one Earth rotation (black circles) with predictions of the spring season approximation, shown in red. Included for comparison is the "EarthObs1" VPL model (dashed blue lines). Panels (a) and (b) show measurements in the "VIO" and "NIR" camera channels. Panels (c) and (d) show measurements by the IR spectrometer at continuum wavelengths between $H_2O$ bands. At 2.2 μm, most of the observed modulation is due to clouds. Models and measurements are in mutual agreement, within uncertainties.*

6.1 Earthshine at High Lunar Latitudes

As an illumination source, earthshine by itself, or in combination with scattered sunlight, should enable optical searches for water frost in PSRs from the surface or from orbit – both important exploration themes. Large areas near both poles have been identified as being permanently shadowed from direct sunlight, while regularly illuminated by Earth as a result of latitude libration (*Mazarico et al., 2011*). The example in Fig. 15 demonstrates the feasibility of high-latitude water frost detection for a single but plausible set of conditions.

Earthshine is reflected from a patch of lunar regolith, and an adjacent patch with frost at 3% fractional area. The sensor is located on the surface with Earth behind it (viewing geometry shown inset in Fig. 15b), so that the observer phase angle, i.e., the angle defined by Earth, target surface and sensor, is close to zero. For Earth's spectral irradiance, we use the diurnal average value at 60° phase angle (Eqn. 4 with $A_L = 0$). The surface is defined by the Hapke rough surface reflectance function (*Hapke 1984, 2012*), extended to near-IR wavelengths, as described by Kennelly et al. (*2010*) based on measurements by the Robotic Lunar Observatory (ROLO). For the dry surface single scattering albedo ($\varpi$) we use the Kennelly et al. (2010) "highlands" values. The single scattering albedo for water ice is obtained using optical constants from Warren and Brandt (2008), and applying the Hapke equivalent-slab model (*Hapke, 2012; Lucey et al., 2014*).

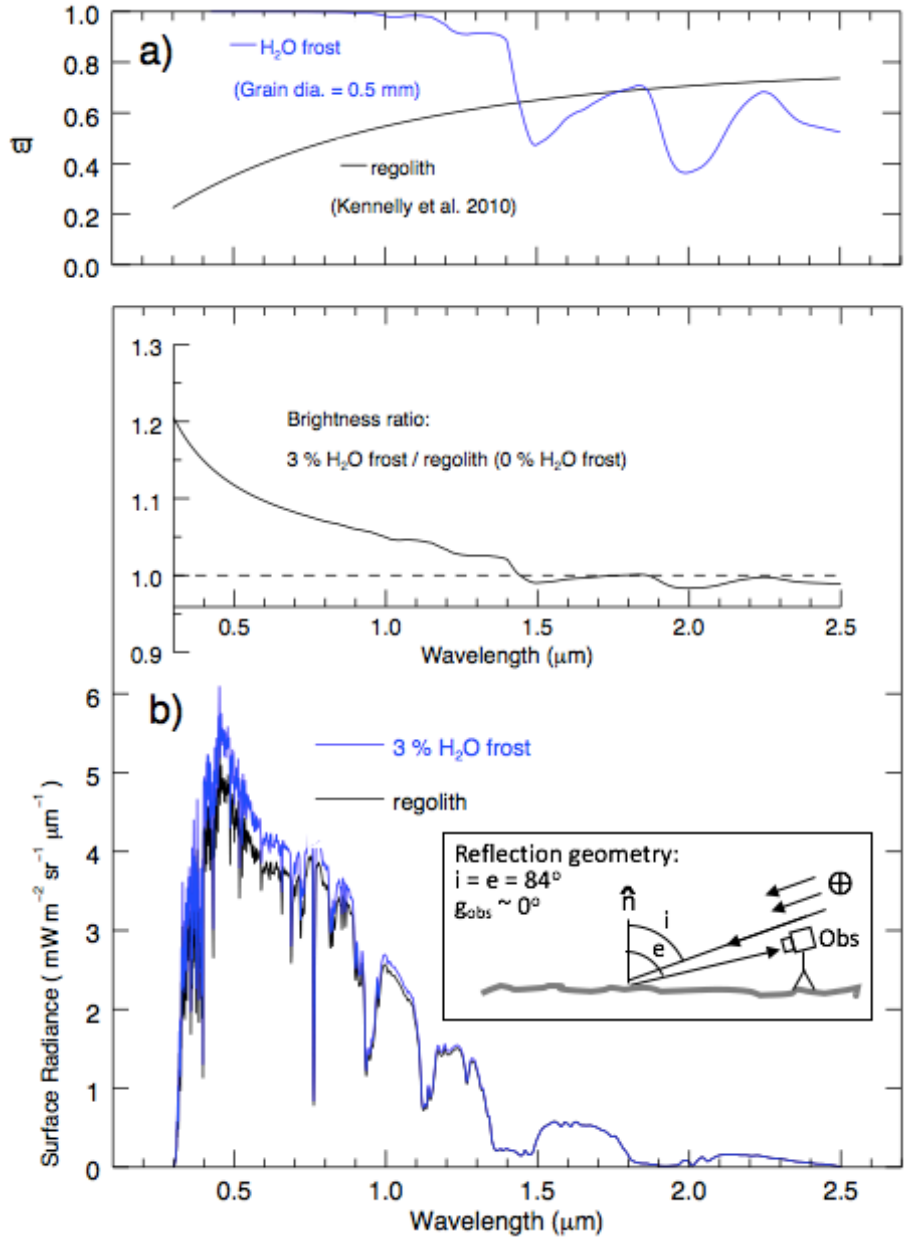

*Figure 15. Simulation of Earthshine reflected from high latitude terrain, with and without a water frost component. (a) Single scattering albedo for dry regolith and water frost consisting of 0.5 mm diameter ice grains. (b) Simulated reflectance, as observed by a camera or spectrometer on the surface. Elevation angles of source (Earth) and detector are 6°, with Earth located behind the sensor (inset). At visible wavelengths, water frost should be detectable using multi-wavelength (e.g. ratiometric) imaging or spectroscopy.*

At visible wavelengths, the intensity of reflected light is bright enough for high SNR, narrow-band measurements with short ($\tau < 1$ s) integration times. The presence of water frost can be readily detected at a fractional coverage of only a few percent by measuring the spectral and spatial contrast between ice-bearing and adjacent ice-free patches as illustrated in Fig.15b. Contrary to this result, detection of water ice at near-IR wavelengths does not appear feasible using earthshine as a light source. This is a consequence of the rapid decline of illumination with increasing wavelength, combined with the presence of deep telluric water vapor bands that are nearly coincident with ice absorption features.

6.2 Limits of Applicability of the Spring Season Approximation

How accurately would the seasonal approximation presented here predict earthshine irradiance in any given year? Interannual changes in Earth's apparent albedo observed over two decades by satellite measurements and terrestrial ground-based measurements of earthshine reflected from the Moon were found to be ~6% in radiance (*Palle et al. 2004, 2009*). This value is comparable to estimates of 5–6% for the variability of full disk reflectance at small phase angles resulting from clouds (*Oakley and Cash, 2009; Robinson et al., 2010*). This would imply that predicting earthshine irradiance as a function of phase is repeatable in subsequent years to well within the overall uncertainties in our empirical approximation.

The amplitude of the diurnal modulation in the March-April 2008 VPL sequence, and therefore in our seasonal approximation, is typically ~30% at near-IR continuum wavelengths, although it is also dependent on the terrestrial sub-observer location and strongly dependent on wavelength. These results presented here are broadly consistent with previous estimates of diurnal brightness change. Palle et al. (*2003*) show a diurnal amplitude of ~25% from model comparisons with ground-based measurements of lunar-reflected earthshine. These results were mean values over a very broad (0.2–4 µm) wavelength band, so a portion of the surface spectrum would be obscured by water vapor absorption in that case. More puzzling is the disagreement with the EPOXI camera measurement at near-IR wavelengths, which show a smaller diurnal amplitude than that predicted by both VPL model and our approximation, as discussed in Sec. 5.3.

Additional approximations of earthshine as seen from the Moon can be computed for other seasons (which we have not done) in order to complete a full-year analysis. Because the Earth-observing geometry is symmetric near equinox, Earth's radiance during other seasons should not be very different when averaged hemispherically and diurnally. To check this assumption, we examined a sequence of VPL Earth models that were created for an extrasolar perspective, hence viewed from fixed longitude in the orbital plane (http://vpl.astro.washington.edu/spectra/planetary/earth_orbit.htm). At small phase angles, Earth is close to summer solstice ($\theta \sim \theta_\odot \sim 23.5°$). Near $g = 0$, we found very little difference between our approximation for the average solar-band radiance and that of the online archive (77 and 80 W m$^{-2}$ sr$^{-1}$ respectively), despite the difference in sub-solar latitude between these simulations. Thermal radiance values were indistinguishable (both 75 W m$^{-2}$ sr$^{-1}$). One would expect greater seasonal dependence at specific wavelengths, but the results of this spring-season study do appear to provide a good estimate of Earth's radiance at any season.

## 7. Summary and Conclusions

We have examined Earth's spectral irradiance at the Moon using 3D models from the NASA Astrobiology Institute's Virtual Planetary Laboratory. The accuracy of VPL Earth radiance predictions has been established by comparison with "EarthObs" measurements during the EPOXI mission.

At most wavelengths, Earth would appear very bright to a lunar observer and could serve as a natural illumination source for future optical measurements of the surface. At solar-reflectance wavelengths (0.3–3.0 µm) the dominant spectral features are strong ozone absorption below $\lambda \sim 0.33$ µm, and an abrupt transition to optically thick (although bright) Rayleigh scattering between 0.33 and ~0.40 µm. At longer wavelengths, the atmosphere becomes mostly transparent, except where the spectrum is marked by strong water vapor absorption bands. Portions of the spectrum inbetween these bands show a strong diurnal intensity modulation, as a result of the changing fraction of observable land area from the major continents and the longitudinal distribution of clouds. At near-IR wavelengths, Earth remains surprisingly bright at high phase angles, as a consequence of ocean glint and the forward-scattering properties of limb clouds.

Near spring equinox, broadband earthshine irradiance ranges from ~60 mW m$^{-2}$ (thermal contribution alone) to as high as 150 mW m$^{-2}$ near zero phase angle. This input is many times the estimate for the Moon's internal heat flow of 9–13 mW m$^{-2}$. It follows that earthshine is a potentially important energy source in models for the long-term stability of volatiles (other than $H_2O$) in the very coldest, high latitude, permanently shadowed terrains.

An approximation for Earth's spring season spectral radiance was created by fitting a simple, semi-empirical functional form to a series of VPL models, using projected land area as a proxy for diurnal brightness modulation. Within its range of applicability ($g < 60°$ and sub-observer latitude between ~30°S and 30°N), this can be used to estimate earthshine irradiance using only ephemeris information. Within the stated uncertainties, we expect that this approximation can be applied to the remaining seasons, and serve as a useful year-round approximation for Earth's spectral irradiance at the Moon.

## Acknowledgements

DAG and TJS were supported by NASA/LASER grant NNX09AO79G and by the NASA Solar System Exploration Virtual Institute SSERVI/DREAM2. TR acknowledges support from the NASA Astrobiology Institute under solicitation NNH12ZDA002C and Cooperative Agreement Number NNA13AA93A, NASA's Sagan Fellowship Program executed by the NASA Exoplanet Science Institute, and NASA's Exoplanets Research Program under Award Number #80NSSC18K0349. EWS also received support from the NASA Postdoctoral Program, administered by the Universities Space Research Association and the NASA Astrobiology Institute, under Cooperative Agreement No. NNA15BB03A.